\documentclass[twocolumn,showpacs,preprintnumbers,amsmath,amssymb,superscriptaddress]{revtex4}
\usepackage{epsfig,bm}
\usepackage{ulem}
\usepackage{amsmath}
\usepackage{color}
\begin{document}
\title{Correlated-basis description of
$\alpha$-cluster and delocalized $0^+$ states in $^{16}$O}
\author{W. Horiuchi}
\affiliation{Department of Physics, Hokkaido University, Sapporo 060-0810, Japan}
\author{Y. Suzuki}
\affiliation{Department of Physics, Niigata University, Niigata 950-2181, Japan}
\affiliation{RIKEN Nishina Center, Wako 351-0198, Japan}

\begin{abstract}
A five-body calculation of $^{12}$C+$n$+$n$+$p$+$p$ is performed to take a step 
towards solving an outstanding problem in nuclear theory:
The simultaneous and accurate description of the  
ground and first excited 0$^+$ states of $^{16}$O.   
The interactions between the constituent particles 
are chosen consistently with the energies of bound subsystems, especially 
$^{12}$C$+n$, $^{12}$C$+p$, and $\alpha$-particle. 
The five-body dynamics is solved with 
the stochastic variational method on correlated Gaussian basis functions.
No restriction is imposed on the four-nucleon configurations 
except the Pauli principle excluding the occupied orbits in $^{12}$C. 
The energies of both the ground and first excited states of $^{16}$O are obtained in excellent agreement with experiment. 
Analysis of the wave functions indicates spatially localized $\alpha$-particle-like 
cluster structure for the excited state and shell-model-like delocalized structure for the ground state.
\end{abstract}
\maketitle

The nucleus $^{16}$O is doubly magic and tightly bound. Its $0^+$ ground 
state is regarded to have predominantly spherical closed shell structure. 
Contradicting the  nuclear shell-model filling 
of single-particle orbits, however, 
the first excited state of $^{16}$O has 
positive parity, $J^{\pi}=0^+$, with the unexpectedly low  
excitation energy $E_x$=6.05\,MeV. 
Its appearance was therefore mysterious.  
A conventional idea is to explain the excited $0^+_2$ state 
with multi-particle multi-hole, especially 4p-4h, excitations. 
The physics mechanism behind such excitations 
is believed to originate from nuclear deformation~\cite{brown66}, and  
the appearance of spherical and deformed states observed 
in several nuclei is called shape coexistence~\cite{rowe06,wood92}. 
The essence of the phenomenon lies in that two states 
with identical quantum numbers are realized at close energies. 
Understanding the coexistence mechanism can thus be a general, interesting problem for other 
quantum many-body systems as well.

Recent theoretical works have focused on the first excited state of $^{16}$O with 
various approaches. Based on the harmonic-oscillator (HO) shell-model 
the possibility of selecting important basis states 
with symplectic algebra~\cite{rowe06} or the modification of 
single-particle energies~\cite{utsuno11} has been discussed.  
Beyond mean-field approaches have been tested in configuration mixing calculation of 
Slater determinants~\cite{bender03,shinohara06}. 
Though the energy gain of the 4p-4h state is found to be substantial   
in the generator coordinate method, its component in the $0^+_2$ state is 
not very large~\cite{bender03}. The basis states in Ref.~\cite{shinohara06} are 
generated by an imaginary-time evolution of stochastically 
selected single-particle Gauss packets, allowing for   
$^{12}$C+$\alpha$-like configurations, 
but the excitation energy of the $0^+_2$ state is too high. 
Large-scale {\it ab initio} calculations with the 
no-core shell model~\cite{maris09} and 
the coupled-cluster theory~\cite{wloch05} have been performed
but the energy of the $0^+_2$ state is still so high in the current
model space that more computational efforts appear to be required 
to reproduce its excitation energy.  
 
It was reported about 40 years ago~\cite{suzuki76} that all $T=0$ levels 
of $^{16}$O but the $10.96\,(0^-)$ state below $E_x=15$\,MeV 
are reproduced by a semi-microscopic $^{12}$C+$\alpha$ two-cluster model where the excitation of $^{12}$C and the Pauli principle are taken into account. 
A microscopic version of the similar cluster model also succeeded  
in reproducing the two $0^+$ states~\cite{descouvemont87}. 
The success seems to suggest that 
the structure of the $0^+_2$ state is closely related to 
the tight binding of $\alpha$-particle, that is, the four particles tend to form an 
$\alpha$-cluster~\cite{horiuchi68}. It should be noted that  
the cluster model space includes some deformation and  for low HO excitations   
has significant overlap with symplectic basis states~\cite{suzuki86}. 

In this paper we report a first converged five-body calculation  
of a $^{12}$C core plus four (valence) nucleons ($4N$) for the $0^+$ states 
of $^{16}$O. This is 
an extension of the work~\cite{suzuki76} towards a more microscopic direction 
in that no preformed $\alpha$-cluster is assumed. 
The excitation of $^{12}$C is ignored. 
Regarding the core as  $0p_{3/2}$ closed configuration,  
we impose the Pauli requirement
that  the valence nucleon be free from the occupied orbits.
Except for that the model has no restriction on the valence nucleon 
orbits, and hence can accommodate 
not only 0p-0h, 2p-2h, 4p-4h, etc. but also $^{12}$C+$\alpha$ configurations.  
To be realistic, both the core-nucleon 
($CN$) and the two-nucleon ($NN$) interactions are chosen consistently with the energies of relevant subsystems, especially  
$^{13}$C ($^{13}$N) and $\alpha$-particle. 
We also treat $^{16}$C as the $^{12}$C core plus 
four neutrons to examine how the $nn$ and $np$ interactions affect the structure.

The five-body system we consider here is described with the following Hamiltonian
\begin{align}
H=T_{v}+T_{cv}+V_v+V_{cv}.
\end{align}
The total kinetic energy  
consists of the kinetic energy of the $4N$ ($T_{v}=\sum_{i=1}^4T_i-T_{c.m.}$) 
relative to their center of 
mass (c.m.) and 
the kinetic energy for the relative motion ($T_{cv}$) between the $4N$ 
c.m. and the core. 
The total potential energy also consists of two terms, $V_v=\sum_{i<j}v_{ij}$ 
and $V_{cv}=\sum_{i=1}^4U_i$. 
The term $v_{ij}$ represents the $NN$ potentials  
between $i$th and $j$th valence nucleons, and
$U_i$ is the $CN$ potential acting on the $i$th nucleon. 
The former is taken from the central Minnesota (MN) potential~\cite{MN}  
that reproduces fairly well the binding energies of $A=2-4$ systems. 
To fine tune the binding energy of $\alpha$-particle, 
the potential strengths of the MN potential are multiplied by 0.9814. 
The latter contains central and spin-orbit terms whose form factors are specified by symmetrized Woods-Saxon (0.65 and 1.25$\times 12^{1/3}$\, fm for the diffuseness and radius parameters) and its derivative, respectively. 
The strength parameters of each term, $V_c^\pi$ and $V_{ls}^\pi$, 
are parity ($\pi$) dependent and set to reproduce the low-lying states of $^{13}$C, $-$4.95 ($1/2^-$),  $-$1.86 ($1/2^+$), and  $-$1.09\,MeV ($5/2^+$) from $^{12}$C+$n$ threshold: $V_c^{-}=-45.78$ MeV, $V_{ls}^{-}=31.08$ MeV\,fm$^2$, 
and $V_{c}^{+}=-57.57$ MeV, $V_{ls}^{+}=17.61$ MeV\,fm$^2$. 
The Coulomb potential is included.

To fulfill the Pauli requirement, a solution $\Psi$ that we want to obtain 
should satisfy the condition 
\begin{align}
\Gamma_i\left|\Psi \right> =0 
\label{ocm}
\end{align}
for $i=1, \dots, 4$, where $\Gamma_i$, acting on the $i$th valence nucleon, is a projector to $0s_{1/2}$ and $0p_{3/2}$ HO orbits 
\begin{align}
\Gamma=\sum_{m}\bigl|0s_{\frac{1}{2}m}\bigr>\bigl<0s_{\frac{1}{2}m}\bigr| 
+\sum_{m}\bigl|0p_{\frac{3}{2}m}\bigr>\bigl<0p_{\frac{3}{2}m}\bigr|,
\end{align}
where $m$ runs over all possible magnetic quantum numbers. 
The radial coordinate of the HO orbit is taken to be the $CN$ relative 
distance vector, and the HO frequency $\hbar \omega$ is set to be 
16.0\,MeV, which reproduces the size of the $^{12}$C ground state.
To practically satisfy the condition~(\ref{ocm}), 
we follow an orthogonality projection method~\cite{kukulin}, in which 
a pseudo potential $\lambda \sum_{i=1}^4\Gamma_i$ with a large 
value of $\lambda$ is added to the Hamiltonian and an energy minimization is carried out. By taking 
$\lambda=10^4$\,MeV, our solution contains vanishingly small 
Pauli-forbidden components of the order of $10^{-4}$. 

The present problem belongs to a class of quantum few-body problems with
orthogonally constraints. This type of problem   
often appears in atomic and subatomic physics 
when the system contains composite particles~\cite{svm}. 
Solving such a problem is quite challenging and 
much effort has been made to eliminate the forbidden states.
Most calculations with the orthogonality constraint have  so far been limited to three- or four-body systems. 
It is only recent that a five-body calculation is performed for 
$^{11}_{\Lambda\Lambda}$Be in the model of 
$\Lambda+\Lambda+\alpha+\alpha+n$~\cite{hiyama10}, 
where the pairwise relative motion of  
$\alpha-\alpha$ and $\alpha-n$ contains Pauli-forbidden states. 
In that hypernuclear case three different relative coordinates 
are involved in the Pauli constraint, while  
in our case the Pauli constraint acts on the four $CN$ coordinates. 
To our knowledge, we here present a first converged solution 
for the core plus four-nucleon five-body system. 

We find a solution by a variational method. 
A trial function has to be flexible enough to satisfy  
several requirements for, e.g., describing different types 
of structure and correlated motion of the particles, 
eliminating the Pauli-forbidden 
components, and accurately describing the tail of the bound-state 
wave function in the asymptotic region.
The trial function is expressed as a combination 
of correlated Gaussian (CG) basis states~\cite{varga95,svm,suzuki08}, 
\begin{align}
\mathcal{A}\left\{{\rm e}^{-\textstyle\frac{1}{2}\tilde{\bm x} A \bm x}\big[[{\cal Y}_{L_1}(\tilde{u}_1\bm x) 
{\cal Y}_{L_2}(\tilde{u}_2\bm x)]_L\chi_{L}\big]\eta_{TM_T}\right\},
\label{cg.basis}
\end{align}
with ${\cal Y}_{\ell}(\bm r)=r^{\ell}Y_{\ell}(\hat{\bm r})$. 
Here ${\cal A}$ is the antisymmetrizer for $4N$, $\bm x$ stands 
for 4 relative coordinates, ($\bm{x}_1,\dots,\bm{x}_4$), 
$A$ is a $4\times 4$ positive-definite, symmetric matrix,  
and $u_1$ and $u_2$ are $4\times 1$ matrices (see \cite{suzuki08} for detail). The elements of $A, u_1, u_2$ as well as $L_1, L_2, L$ 
are continuous and discrete variational parameters, respectively. 
The function $\chi$ ($\eta $) specifies spin (isospin) 
states of $4N$. Possible $L$ values are 0, 1, and 2. The c.m. motion of the total system 
is excluded  in Eq.~(\ref{cg.basis}), and no spurious c.m. motion is included. 

The power of the CG basis of type~(\ref{cg.basis}) has been demonstrated by 
many examples~\cite{suzuki08,horiuchi08,mitroy13}. 
An advantage of the CG is that it keeps its functional form 
under a linear transformation of the coordinates~\cite{varga95,svm}, 
which is a key for describing  
both cluster and delocalized structure in a unified manner. 
Each basis element contains so many variational parameters 
that discretizing them on grids 
leads to an enormous dimension ${\cal K}$ of at least $10^{10}$. 
Thus we test a number of candidate bases with the stochastic variational 
method~\cite{kukulin77,varga94,varga95,svm}, choose the best one among them and increase the basis dimension one by one until 
a convergence is reached. This procedure costs expensively for computer time but no other 
viable methods are at hand to get converged solutions for the present problem.  

\begin{figure}[t]
\begin{center}
\epsfig{file=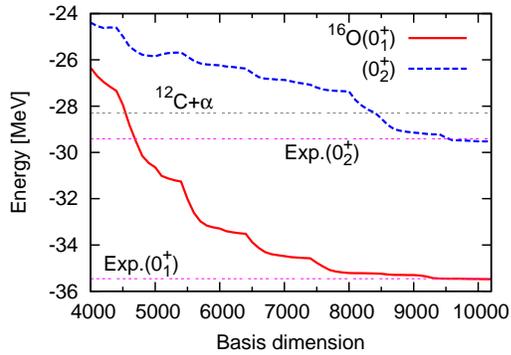,scale=1.1}
\caption{(Color online) Energies from $^{12}$C+$n$+$n$+$p$+$p$ threshold for 
the ground and first excited $0^+$ states of $^{16}$O.
The  $^{12}$C$+\alpha$ threshold  
and experimental energies are shown by thin lines.}
\label{conv16O.fig}
\end{center}
\end{figure}

Figure~\ref{conv16O.fig} displays the energies of 
two lowest $0^+$  states of $^{16}$O versus the 
basis dimension.  More than 9,500 bases are combined to reach the convergence. 
Most bases, especially up to ${\cal K}=4,000$, first 
serve to eliminate the forbidden states, which is because the use of 
large $\lambda$ value to ensure the Pauli principle leads to 
large positive energies at small basis dimension.  
The valence nucleons tend to move around the core 
to gain $V_{cv}$ and at the same time they want to correlate among them  
to make use of the attraction of $V_v$. 
Eliminating the forbidden states under such competition is  hard. 
After the ground state energy converges well,  
the variational parameters are searched to optimize the first excited state 
at ${\cal K}=8,000-9,000$. The energy gain after 
${\cal K}=9,500$ is very small.    
Two $0^+$ states appear below $^{12}$C+$\alpha$ threshold and their 
energies are both remarkably close to experiment.  
Compared to $^{16}$O, the convergence for $^{16}$C is 
faster: 7,000 bases are enough to describe the weaker correlated motion 
of $4N$ and reproduce the ground state energy very well.
The obtained energies are listed in Table~\ref{edecom.tab}. 
We repeated the calculation with the original MN potential. 
The binding energy of 
$\alpha$-particle increased by about 1\,MeV, but the energies of the 
two $0^+$ states from the threshold virtually remained unchanged. 

\begin{table}[b]
\begin{center}
\caption{ Energy contents in MeV and root-mean-square (rms) radii in fm 
of the $0^+$ states of $^{16}$O and $^{16}$C.   
The results of $\alpha$-particle are due to a  
four-body calculation with the MN potential.
Empirical rms radii are taken from \cite{ozawa01, zheng02} 
for $^{16}$C and \cite{angelia13} for $\alpha$ and $^{16}$O.}
\begin{tabular}{ccccccccc}
\hline\hline
   &&$^{16}$C ($0^+_1$)&&$^{16}$O ($0^+_1$)&&$^{16}$O ($0^+_2$)&&$\alpha$\\
\hline
$E$&&$-$18.47&&$-$35.47&&$-$29.52 &&$-$28.30\\
$E_{\rm exp.}$&&$-$18.59&&$-$35.46&&$-$29.41 &&$-$28.30\\
$\left<T_{cv}\right>$&& 17.81&& 11.55&&7.16 &&--\\
$\left<V_{cv}\right>$&&$-$82.49&&$-$79.55&&$-$29.22&&--\\
$\left<T_v\right>$&&53.53 && 72.93&& 67.46&&56.92\\
$\left<V_v\right>$&&$-$7.32 &&$-$40.41 &&$-$74.92&&$-$85.22\\
\hline
$\sqrt{\langle r^2\rangle}$ &&2.62 &&2.47 &&3.03&&1.43\\
$\sqrt{\langle r^2\rangle}_{\rm exp.}$ &&2.70(3), 2.64(5) &&2.57(2) &&--&&1.46(1) \\
$\sqrt{\langle r_{cv}^2\rangle}$&&1.94&& 2.54&&4.86&&--\\
$\sqrt{\langle r_{v}^2 \rangle}$&&2.88&& 1.90&&1.62&&1.43\\
\hline\hline
\end{tabular}
\label{edecom.tab}
\end{center}
\end{table}

Analyzing the contribution of each piece of the Hamiltonian to the energy  is 
important to understand the binding mechanism.  
As listed in Table~\ref{edecom.tab}, in $^{16}$C the attraction mainly comes from $V_{cv}$.   
In the ground state of $^{16}$O, similarly to $^{16}$C,   
$V_{cv}$ is still a major source of the attraction 
but $V_v$ also contributes to the energy significantly, 
which should not come as a surprise given that 
the $np$ interaction is more attractive than the $nn$ interaction.  
Since $\left<V_v\right>$ is about  
a half of that of $\alpha$-particle, the $4N$ in the ground state of $^{16}$O are 
strongly distorted from the intrinsic state of $\alpha$-particle due to 
both the $CN$ interaction and the Pauli constraint. 
The first excited state of $^{16}$O exhibits an opposite pattern. 
The contribution of $V_v$ is dominating and close 
to that of $\alpha$-particle. It looks that the first excited state has 
$^{12}$C$+\alpha$ cluster structure 
as shown by the cluster model~\cite{suzuki76}. We note, however, that the 
$4N$ in the $0^+_2$ state are not as strongly bound as $\alpha$-particle. 
In fact the $4N$ internal energy, 
$\left<T_v\right>$+$\left<V_v\right>$, is only about a quarter of 
that of $\alpha$-particle. 
The two $0^+$ states of $^{16}$O have a different face but coexist 
closely in energy due to the combined function of the $NN$ and $CN$ interactions.  

\begin{figure}[t]
\begin{center}
\epsfig{file=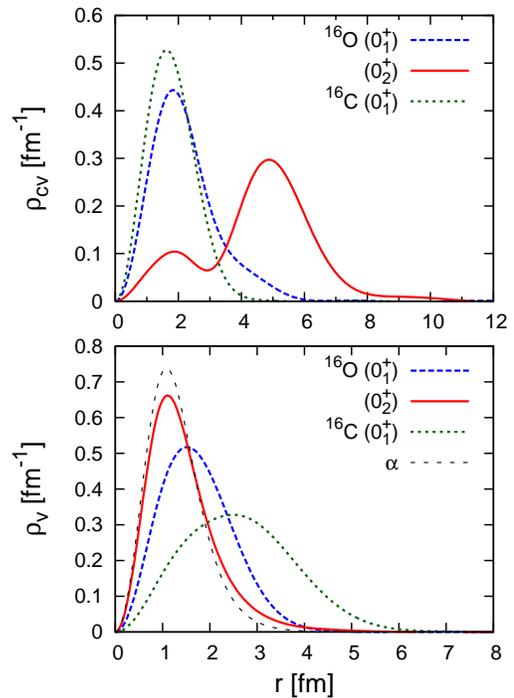,scale=1.1}
\caption{(Color online). Top: Distributions of the relative distance  
between the $^{12}$C core and the c.m. of $4N$. 
Bottom: Density distributions of the valence nucleon measured
from the $4N$ c.m.
The nucleon density distribution of $\alpha$-particle is calculated using the MN potential.}
\label{dens16.fig}
\end{center}
\end{figure}

The different structure discussed above is visualized by 
comparing the spatial properties of the three states. 
Top panel of Fig.~\ref{dens16.fig} shows 
$4N$ c.m.-core relative motion distribution,  
$\rho_{cv}(r)=\langle \delta(|\bm r_v-\bm r_c|-r) \rangle$, 
where $\bm r_v$ and $\bm r_c$ are the coordinates of 
the $4N$ c.m. and the core, and bottom one 
the valence nucleon distribution in $4N$, 
$\rho_{v}(r)=\langle \delta(|\bm r_1-\bm r_v|-r) \rangle$. 
In case of $^{16}$C, $\rho_{cv}$ is narrow whereas   
$\rho_{v}$ is spread. 
Four neutrons move on certain orbits with small radii while  
being apart from each other, indicating an independent particle like motion.
In contrast to $^{16}$C, the $0_2^+$ state of $^{16}$O shows not only extended 
$\rho_{cv}$ whose highest peak is at about $^{12}$C+$\alpha$ touching distance ($\sim$4.9\,fm),  
but also such narrow $\rho_{v}$ that is very similar to 
the density distribution of $\alpha$-particle. This supports 
that the $0_2^+$ state of $^{16}$O 
has $\alpha$-cluster-like structure.  The distribution of the ground state of $^{16}$O is somewhat 
intermediate between $^{16}$C and the $0_2^+$ state of $^{16}$O. 

The rms radii are listed in Table~\ref{edecom.tab}, where,  
e.g., $\langle r_{cv}^2\rangle$ stands for $\int_0^{\infty} r^2 \rho_{cv}(r)dr$. 
The point matter radius, $\sqrt{\langle r^2\rangle}$, is obtained assuming 
the rms radius of $^{12}$C core as 2.33\,fm~\cite{angelia13}. 
The matter radii for the ground states of $^{16}$C and $^{16}$O agree with experiment fairly well. 
Supporting the $\alpha$-cluster structure, $\sqrt{\langle r_{cv}^2\rangle}$  
of the $^{16}$O($0_2^+$) state is two times larger than that of the $^{16}$O ground state, while  
$\sqrt{\langle r_{v}^2 \rangle}$ is small and slightly larger than the  radius 
of $\alpha$-particle. 
The ratio  $\gamma=\sqrt{\langle r_{cv}^2\rangle}/\sqrt{\langle r_{v}^2 \rangle}$ may serve as a 
measure of clustering. The larger $\gamma$, the more prominent the clustering. The $\gamma$ value is   
0.67 for $^{16}$C and grows to 1.3 and 3.0 for the ground and excited states of $^{16}$O.    
The monopole matrix element, $|\left<0_2^+\right|r_p^2\left|0_1^+\right>|$, for the two $0^+$ states of $^{16}$O is 6.55\,fm$^2$, somewhat larger than 
experiment (3.55$\pm$0.21\,fm$^2$~\cite{tilley93}), which may be improved 
by allowing for the excitation of $^{12}$C core. 

\begin{figure}[b]
\begin{center}
\epsfig{file=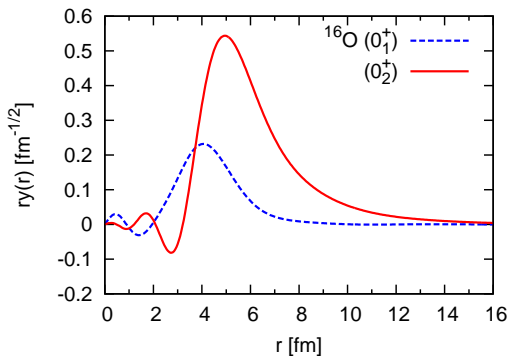,scale=1.1}
\caption{(Color online). 
$^{12}$C$+\alpha$ spectroscopic amplitudes for the ground and
first excited $0^+$ states of $^{16}$O.}
\label{spectro.fig}
\end{center}
\end{figure}

As a measure of finding $\alpha$-particle as a function of the distance $|\bm r_v-\bm r_c|$,  
we draw in Fig.~\ref{spectro.fig}  
$^{12}$C$+\alpha$ spectroscopic amplitudes for the two $0^+$ states, 
\begin{align}
y(r)=\frac{1}{r^2}\Bigl<\phi_\alpha\,\delta (|\bm r_v-\bm r_c| - r)Y_{00}(\widehat{\bm r_v-\bm r_c})\Bigl|\Bigr.
\Psi\Bigr>,
\end{align}
where $\phi_\alpha$ is the $\alpha$-particle wave function 
obtained with the MN potential. Two curves show a striking difference.  
In the $0_1^+$ state, the highest peak is located near the surface region of 
the core. 
The spectroscopic factor, $S_{\alpha}=\int_0^\infty [ry(r)]^2 dr$,
is small (0.105). 
Compared to this, the amplitude of the $0_2^+$ state 
is much larger and has a peak at  
$^{12}$C+$\alpha$ touching distance. It is by far larger and longer ranged than that calculated by the deformed model~\cite{ichimura73}. 
The $S_{\alpha}$ value is 0.680, in agreement with 0.679 of 
the $^{12}$C$+\alpha$ cluster model~\cite{suzuki76}. 
The dimensionless reduced 
$\alpha$-width, $\theta_{\alpha}^2$, at a channel radius $r$, 
defined by $r^3[y(r)]^2/3$, is a 
better measure of $\alpha$ clustering than $S_{\alpha}$. 
The value is 0.341 at $r=6$\,fm, large enough 
to be compared to that of the negative-parity rotational 
band based on the $9.59 \,(1^-_2)$ state of 
$^{16}$O~\cite{suzuki76}. 

The behavior of  $\rho_{cv}(r)$ and $y(r)$ shown in 
Figs.~\ref{dens16.fig} and ~\ref{spectro.fig} is understood as follows.
Letting $\bm{x}_4$ denote $\bm{r}_v-\bm{r}_c$, we may write those functions as 
\begin{align}
y(r)&\propto \int\,d\hat{\bm{r}}\int
d\bm{x}_{v}\,\phi_\alpha^*(\bm{x}_{v})\Psi(\bm{x}_{v},\bm{r}),\\
\rho_{cv}(r)&\propto r^2 \int d\hat{\bm{r}}\,\int d\bm{x}_{v} 
\left|\Psi(\bm{x}_{v},\bm{r})\right|^2,
\label{def.rho}
\end{align}
where $\bm{x}_v$ collectively stands for 3 internal coordinates of the valence nucleons 
and the spin and isospin coordinates as well as the relevant integration over those coordinates are abbreviated. First we discuss $y(r)$. As shown in Fig.~\ref{spectro.fig},
the spectroscopic amplitudes for both the ground and first excited states 
are suppressed and exhibit nodal behavior at short distances. This is 
because  $\Psi$ contains no $0s_{1/2}$ and $0p_{3/2}$ orbits owing to 
the Pauli principle and $y(r)$ contains at least 4$\hbar\omega$ HO components.
Next we discuss $\rho_{cv}(r)$. It is clear from Eq.~(\ref{def.rho}) that $\rho_{cv}(r)$ 
is non-negative and its behavior at small $r$ is determined mainly by those orbits 
that have relatively small radii such as $0p_{1/2}, 1s_{1/2}$, etc. The lower  
bump of $\rho_{cv}(r)$ for the excited state is a consequence of the fact 
that the wave function of the excited state is orthogonal to 
the ground state wave function.

\begin{figure}[t]
\begin{center}
\epsfig{file=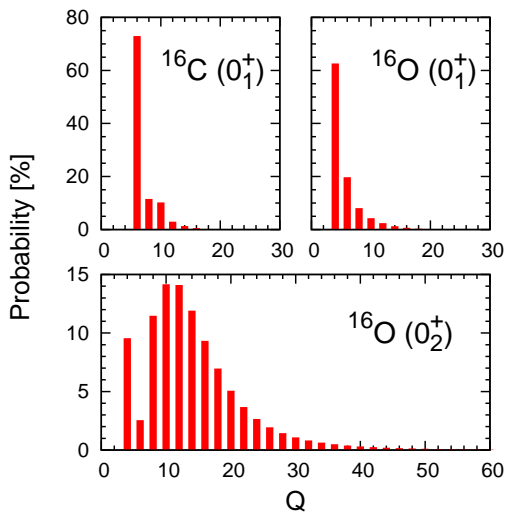,scale=1.1}
\caption{(Color online). Decomposition of the $0^+$ states of $^{16}$O and $^{16}$C 
into $Q\hbar \omega$ components with $\hbar \omega=16.0$\,MeV.}
\label{HOquanta.fig}
\end{center}
\end{figure}

It is useful to expand the obtained wave functions in terms of the HO basis, 
especially because the $0^+_2$ state challenges no-core  
shell-model description~\cite{maris09}. 
Explicit expansion is not feasible but counting the number of HO quanta
is easy~\cite{suzuki96}. 
Figure~\ref{HOquanta.fig} plots the probability 
of $Q\hbar \omega$ components occupied by $4N$. 
The distribution for $^{16}$C and $^{16}$O($0^+_1$) is normal: 
The largest probability occurs at minimum $Q$ 
(6 for $^{16}$C and 4 for $^{16}$O) 
and decreases rapidly with increasing $Q$. 
The average ($M_Q$) and standard deviation ($\sigma_Q$) of $Q$-distribution is  
7.0 and 2.1 for $^{16}$C, and 5.5 and 2.9 for the ground state of $^{16}$O, 
respectively. 
In contrast with this normal case, the distribution for the excited state of 
$^{16}$O exhibits a quite different pattern. The probability 
is widely distributed 
and not negligible even beyond $Q=20$, with $M_Q=14.3$ and $\sigma_Q=8.3$. 
The peak at $Q=10-12$ corresponds to $2-4\hbar \omega$ 
more excitation than 4p-4h.  
A distribution similar to the $0^+_2$ case is also obtained for the Hoyle state ~\cite{neff09,suzuki96}. 
Approach like Monte Carlo shell model~\cite{shimizu13} or no-core shell model with symmetry 
adaptation~\cite{dytrych07}, importance truncation~\cite{roth07}, etc. may be able to 
describe these states in future but developing an innovative method of calculation   
will be indispensable. 

The core excitation is ignored in the present study. 
If we allow for the core excitation, we first need to construct 
the wave functions of both the ground and excited states of $^{12}$C 
in a microscopic model, and then define the Pauli-forbidden states 
using those wave functions. In addition, the $CN$ potential has to 
be determined consistently with this extended model. 
According to the $^{12}$C+$\alpha$ cluster model calculation~\cite{suzuki76}, 
the core excitation can be ignored in the first excited state of $^{16}$O 
but a certain amount of the excited component is contained in its ground state. 
This is natural because the core excitation occurs more likely 
as the valence nucleon gets closer to the core 
and because the probability of finding the valence nucleon near the core 
is expected to be much larger in the ground state. 
If that is the case, in the ground state the energy loss due to the core excitation 
has to be compensated by some additional attraction of the $CN$ potential. 
Thus the consequence of the core excitation will result in shifting the 
ground state of $^{16}$O towards more delocalized structure. 
Compared to the case with no core excitation, we speculate that 
the peak position of the spectroscopic amplitude gets closer to the core
and the distribution of HO quanta is concentrated more in low oscillator quanta.
This possible change of the ground state structure also helps 
to reduce the monopole strength. 
Further study along this direction is certainly important 
for a more detailed description of the shape coexistence in $^{16}$O.

We have attempted to describe simultaneously 
both the ground and first excited $0^+$ states of $^{16}$O 
in the five-body approach of $^{12}$C plus four nucleons.
The model space is large enough to describe the multi-particle multi-hole excitations, the shape 
coexistence and the $^{12}$C+$\alpha$ clustering. 
Once the potentials between the particles 
are chosen to reproduce the energies of the 
relevant subsystems, neither adjustable parameter 
nor a bias for the existence of $\alpha$-cluster 
is necessary. 
The converged solutions for the five-body Sch\"odinger equation 
with the Pauli constraint are obtained 
with the stochastic variational method 
on the correlated Gaussian basis functions.
The ground state of $^{16}$C treated as the system of $^{12}$C plus four neutrons 
is also examined. 

The energies of the ground and first excited states of $^{16}$O  
as well as the ground state of $^{16}$C  
are all obtained in very good agreement with experiment. 
To understand the coexistence mechanism 
for the two $0^+$ states of $^{16}$O, we analyze the role of 
both the core-nucleon and nucleon-nucleon potentials. In the $0^+_2$ state 
the four nucleons contribute to gaining energy significantly, suggesting 
the formation of $\alpha$-cluster. The different character of the 
two states is exhibited by comparing  
density distributions, particle distances, 
$^{12}$C+$\alpha$ spectroscopic amplitudes, 
and probability distributions of harmonic-oscillator quanta. They all exhibit something like 
a phase transition occurring between delocalized  and cluster structure. 

As further investigation, it is interesting to include the effect of $^{12}$C core excitation on the spectrum of $^{16}$O.  
Extending the present approach to 
heavier nuclei such as $^{20}$Ne, $^{40}$Ca, $^{44,52}$Ti, and $^{212}$Po 
will also be 
interesting for 
exploring a possible universal role of $\alpha$-like correlation in 
shape coexistence and $\alpha$-decay with an increasing mass number of the core nucleus. 

This work was supported in part by JSPS KAKENHI Grant Numbers
21540261, 24540261, and 25800121.

\end{document}